\documentclass[envcountsame]{LMCS}

\usepackage{a4wide}
\usepackage[curve]{xypic}
\usepackage{latexsym}
\input xy
\xyoption{all}
\usepackage{xcolor}\SelectTips{cm}{10}
\usepackage{hyperref}

\usepackage{amsmath}
\newcommand{\red}[1]{\textcolor{black}{#1}}
\newcommand{\blue}[1]{\textcolor{black}{#1}}

\mathcode`<="4268  
\mathcode`>="5269  
\mathcode`;="003B  

\mathchardef\ls="213C    
\mathchardef\gr="213E    

\newtheorem{theorem}[thm]{Theorem}

\newtheorem{definition}[thm]{Definition}
\newtheorem{example}[thm]{Example}
\newtheorem{proposition}[thm]{Proposition}
\newtheorem{corollary}[thm]{Corollary}
\newtheorem{remark}[thm]{Remark}

\newcommand{\comment}[1]{\hspace{1em}\mbox{[{\small #1}]}}
\def\qed{\ifmmode
         $\Box$
         \else
         {\unskip
          \nobreak
          \hfil
          \penalty50
          \hskip1em
          \null
          \nobreak
          \hfil
          $\Box$
          \parfillskip=0pt
          \finalhyphendemerits=0
          \endgraf}
         \fi}

\def\ofR{{\rm I\makebox[- 0.15 em]{}R}}        

    \newcommand{\overto}[1]{\stackrel{\scriptsize{#1}}{%
    \overrightarrow{\smash{\,\scriptsize{\phantom{#1}}\,} }}}

\newbox\aqq  
\newbox\bqq
\newtoks\qaq 
\newtoks\qbq 
\newdimen\lqw

\def\sqtqp#1#2{\mathsurround=0 pt%
  \csname qaq\endcsname={@#1}
  \setbox\aqq=\hbox{$_{\quad #2}$}\lqw=\wd\aqq}
\def\mqr#1#2#3{\sqtqp{#1}{#3}\objectmargin{1 pt}%
  \xy
    (0,0)*+{\mathstrut\kern2 pt}
    \expandafter\ar\the\qaq^-{\raise #2 pt\hbox{$\smash{\scriptstyle #3}$}}
    <\lqw,0 pt>*+{\kern2 pt\mathstrut}
  \endxy\relax}
\def\overto#1{\mqr{{->}}{0}{#1}}

\def\doi{4 (3:9) 2008}
\lmcsheading%
{\doi}
{1--22}
{}
{}
{Dec.~20, 2007}
{Sep.~19, 2008}
{}   

\begin{document}

\title{Rational Streams Coalgebraically}

\author[J.~Rutten]{Jan Rutten}

\address{CWI and VUA, Kruislaan 413, 1098 SJ  Amsterdam}
\email{janr@cwi.nl}

\date{}


\titlecomment{This research is partially supported by
EU project IST-33826 CREDO (\texttt{http://credo.cwi.nl}).}


\begin{abstract}
\noindent
We study {\em rational} streams (over a field)
from a coalgebraic perspective. Exploiting the
finality of the set of streams, we present
an elementary and uniform proof of the equivalence of four
notions of representability of rational streams:
by finite dimensional linear systems; by finite stream circuits;
by finite weighted stream automata;
and by finite dimensional subsystems of the set of streams.
\end{abstract}

\maketitle

\section{Introduction}
\label{Introduction}

A stream over a given set $A$ is an infinite sequence of elements of
$A$. Streams are abound in both mathematics and computer science.
Think of limits in mathematics, typically defined
in terms of converging sequences, and of Taylor series of
analytical functions. In computer science,
streams occur in various fields such as data flow, infinite data types,
semantics, formal power series, and functional programming.

In this paper,
we study {\em rational} streams (over a field).
They are well-known
in mathematics, notably system theory, but have not
received much attention in computer science.
In contrast, a basic ingredient in any introductory course
in the theory of computation is the notion of
rational {\em language}
(also called regular language). Rational languages
are a prototypical example
of a finitely presentable data type: a language is rational
if and only if
it is recognisable by a finite automaton.

As we shall see, rational streams are similarly finitely representable,
in various ways. More specifically, a stream is rational
iff it satisfies one of the following equivalent conditions:
\begin{enumerate}
\item[(a)]
it is representable by a finite dimensional linear system;
\item[(b)]
it is computable by a finite stream circuit;
\item[(c)]
it is representable by a finite weighted stream automaton.
\end{enumerate}
We shall explain the details of all of this as we will go along,
but for now it is worthwhile pointing out that condition (c)
is similar, in the case of languages, to being recognisable by
a finite automaton. Condition (b) is particularly nice
and relevant for computer scientists, since it provides
a very elementary characterisation of rational streams
in terms of finite memory (registers) and feedback.

Streams are for the theory of coalgebra,
one could say, what the natural
numbers are for algebra: a canonical example illustrating
some of the essential notions of the theory.
The set of natural numbers
is an {\em initial} algebra (and satisfies a principle of
{\em induction}). Dually, the set of streams is a {\em final}
coalgebra (and satisfies a principle of {\em coinduction}).
In the present paper, the proofs of the equivalence
of the above three conditions will in essence
be based on the finality of the set of streams.
Finality moreover provides yet another equivalent characterisation
of rationality. A stream is rational
if and only if
\begin{enumerate}
\item[(d)] it generates a finite dimensional subsystem of the set of
all streams.
\end{enumerate}
As we shall see, this criterion is particularly useful for proving
that a stream is {\em not} rational.

Most of the above and of the contents of this paper
is already known, but often at different places in the literature
and typically formulated in different languages, as it were.
The equivalence of rationality and condition (a) above is taken
as a definition in the theory of formal power series
\cite{BR88}.
The equivalence of rationality
and condition (b) is proved in system theory and
in signal processing, where stream circuits are known under various
names (such as signal flow graphs)
\cite{Kai80,Lah98};
see also our earlier paper \cite{Rut07}. The equivalence between
(a) and (c) is known in automata theory.
Condition (d) occurs in some of our own work
\cite{Rut05b};
its use here to disprove rationality of a stream seems
to be new.

All in all,
our main goal has been to present rational streams and all of
their characterisations in one place and in
elementary and uniform terms. Moreover, this paper is intended
as a form of publicity for rational streams to the
computer science community. They provide a basic and simple
model of finite memory and feedback and deserve, therewith,
a place in the heart of the foundations of the theory of computing.
Finally, our treatment of rational streams serves as a good
illustration of the relevance of the combined use
of both algebraic and coalgebraic methods in computer science.

\section{Rational streams}
\label{Rational streams}

We define the set of {\em streams\/} over a given set $A$
by
$$
A^\omega = \{ \sigma \mid \sigma : \{0,1,2, \ldots \} \to A \}
$$
We will denote elements $\sigma \in A^\omega$ by
$\sigma = (\sigma(0), \sigma(1), \sigma(2), \ldots)$.
We define the {\em stream derivative\/} of a stream $\sigma$ by
$$
\sigma' = (\sigma(1), \sigma(2), \sigma(3), \ldots)
$$
and we call $\sigma(0)$ the {\em initial value\/} of $\sigma$.
For $a \in A$ and $\sigma \in A^\omega$ we use the following
notation:
$$
a: \sigma = (a, \sigma(0), \sigma(1), \sigma(2), \ldots )
$$
For instance, $\sigma = \sigma(0) : \sigma'$, for any
$\sigma \in A^\omega$.
(In computer science, the operations of initial value and
derivative are known as {\em head\/} and {\em tail}.)

If the set $A$ carries some algebraic structure
then typically this induces some structure on
the set $A^\omega$ as well.
In particular, if
the set $A$ is a {\em (semi-)ring\/}
\[
A=(A,\,+,\,\cdot \, , \, 0,\,1)
\]
(see the Appendix for the full definition)
then the set $A^\omega$
of streams over $A$ can be equipped with
operations and constants that allow the formulation
of an elementary but useful calculus.

So let $A$ be a (semi-)ring. Examples are the set of real numbers
(which is also a field) and the set of
linear transformations on a vector space.
We define the following operators
on the set $A^\omega$ of streams over $A$,
for all $c \in A$, $\sigma,\tau \in A^\omega$, $n \geq 0$:
\begin{eqnarray*}
[c] &=& (c,0,0,0, \ldots) \;\;\;\;
\mbox{(often simply denoted again by $c$)}
\\
X & = & (0,1,0,0,0, \ldots)
\\
(\sigma + \tau)(n) & = &
\sigma (n) + \tau(n)
\;\;\;\;\;\;
\comment{sum}
\\
(\sigma \times \tau)(n) & = &
\sum_{i=0}^{n}
\sigma (i) \cdot \tau(n-i)
\;\;\;\;\;\;
\comment{convolution product}
\end{eqnarray*}
(where $\; \cdot \;$ denotes multiplication in the ring $A$).
For the above definitions, it is already sufficient if $A$ is
a {\em semi}-ring. If $A$ is moreover a ring then it comes equipped
with an additive inverse, which extends to streams,
for $\sigma \in A^\omega$, as follows:
\[
- \sigma = ( -\sigma(0) , \,  -\sigma(1) , \, -\sigma(2) , \, \ldots )
\]
(here the minus symbols on the right are from the ring $A$).
If the initial value $\sigma(0)$ of a stream $\sigma$ has
a multiplicative inverse $\sigma(0)^{-1}$ in $A$ then
$\sigma$ has a (unique) multiplicative inverse
$\sigma^{-1}$ in $A^\omega$:
\[
\sigma^{-1} \times \sigma = [1]
\]
As usual, we shall often write
$1/\sigma$ for $\sigma^{-1}$. If $\cdot$ and hence $\times$ is commutative,
then we also write
$\sigma / \tau$ for $\sigma \times \tau^{-1}$.

In general, the $n$th element $\sigma(n)$ of a stream
$\sigma$ is trivially given by
\[
\sigma(n) = \sigma^{(n)} (0)
\]
where the superscript $(n)$ denotes the $n$th stream derivative.
In \cite{Rut05b}, various rules for the computation of stream derivatives
are given. For the examples used in the present paper,
all we shall be needing is the very simple rule
presented in Corollary \ref{calculating derivatives}
below.

The stream operators introduced above are well behaved
in that they inherit the properties of the operators
of the underlying (semi-)ring. Notably
sum and product are associative; sum is commutative
but product $\times $ is only commutative if $\cdot$ is;
$[0]$ is the additive identity, $[1]$ is the multiplicative
identity. Another property we shall be using is the following.
For all $\sigma \in A^\omega$,
\begin{equation}
\label{multiplying with X}
X \times \sigma = \, \sigma \times X
\end{equation}
Note that this property also holds for (semi-)rings $A$
in which multiplication is not commutative.

%
Since $X^2 = (0,0,1,0,0,0, \ldots)$,
$X^3 = (0,0,0,1,0,0,0, \ldots)$ and so on,
the following infinite sum is well defined,
for all $\sigma \in A^\omega$:
$$
\sigma = \sigma(0) + (\sigma(1) \times X) +
(\sigma(2) \times X^2 ) +  \cdots
$$
(Note that we write $\sigma(i)$ for
$[\sigma(i)]$; similarly below.)
It shows that $\sigma$ can be viewed as
a formal power series in the indeterminate $X$
(which here in fact is a constant stream).
What distinguishes our approach from
the classical theory of
formal power series is a systematic use
of the operation of stream derivative
and the universal property of finality it induces (cf. Section
\ref{Linear representations}).
This leads to a somewhat non-standard algebraic calculus,
which we call {\em stream calculus\/}.

The following identity shows how one can compute a stream from
its initial value and its derivative. Since this amounts
to a form of (stream) {\em integration} it is called
the {\em fundamental theorem of stream calculus} \cite{Rut05b}.

\begin{thm}
[Fundamental theorem]
\label{fundamental theorem}
{\rm
Let $A$ be a (semi-)ring.
For all $\sigma \in A^\omega$,
\[
\sigma
= \,
\sigma(0) + (X \times \sigma')
\]
}
\end{thm}

\begin{proof}
Immediate from the fact that
$X \times \sigma' = (0,\, \sigma(1) , \, \sigma(2) , \, \sigma(3) , \,
\ldots )$.
\end{proof}

\medskip
\noindent
For future reference, we list the following identities on
initial values, which are immediate from the definition
of the operations on streams.

\begin{proposition}[Initial values]
\label{basic properties of stream calculus}
{\rm
For all $\sigma \in A^\omega$,
\begin{eqnarray*}
(\sigma + \tau)(0)
& = &
\sigma(0) + \tau(0)
\\
(\sigma \times  \tau)(0)
& = &
\sigma(0) \cdot \tau(0)
\\
\sigma^{-1}(0)
& = &
\sigma(0)^{-1}
\end{eqnarray*}
where in the last identity $\sigma(0)$ is assumed
to have a multiplicative inverse in $A$.
\qed
}
\end{proposition}

\medskip
\noindent
Next we introduce the notion of rational stream.

\begin{defi}[Rational streams]
\label{rational streams}
{\rm
We call a stream $\pi$
{\em polynomial\/} if it is of the form
\begin{eqnarray*}
\pi
& = &
c_0 + (c_1 \times X) + (c_2 \times X^2) + \cdots + (c_k \times X^k)
\\
& = &
(c_0,\, c_1 ,\, c_2 ,\, \ldots ,\, c_k , \, 0, \, 0, \, 0, \, \ldots )
\end{eqnarray*}
A stream $\rho$ is {\em rational\/} if it
is the quotient
\[
\rho
= \;
\sigma/\tau
= \;
\sigma \times \tau^{-1}
\]
of two polynomial streams $\sigma$ and $\tau$
for which $\tau(0)^{-1}$ exists.
We denote the set of all rational streams over $A$ by
$$
Rat(A^\omega) =
\{
\sigma \in A^\omega \mid \mbox{$\sigma$ is rational} \,
\}
$$
\qed
}
\end{defi}

\begin{remark}
In the literature, one also encounters the notion
of rational stream defined as being {\em ultimately periodic}.
In the present setting, these streams can be simply characterized
as having only finitely many distinct derivatives.
As we shall see in Section \ref{Constructing linear representations},
it follows from this
basic observation that ultimately periodic streams
are a special case of rational streams in our sense.
\qed
\end{remark}

\medskip
\noindent
Theorem \ref{fundamental theorem}
(together with Proposition \ref{basic properties of stream calculus})
gives an easy calculation rule for the computation
of stream derivatives. First note that
for all $\sigma \in A^\omega$,
\begin{equation}
\label{derivative of X times sigma}
(X \times \sigma )' = \, \sigma
\end{equation}
Furthermore we have, for any $\sigma \in A^\omega$, that
$X \times \sigma ' = \sigma - \sigma(0)$,
by Theorem \ref{fundamental theorem},
and $( X \times \sigma ' ) ' =  \sigma '$,
by (\ref{derivative of X times sigma}).
As a consequence, we have the following.

\begin{corollary}
\label{calculating derivatives}
For all $\sigma \in A^\omega$,
\[
\sigma' = \; ( \sigma - \sigma(0) )'
\]
\qed
\end{corollary}

\medskip
\noindent
This trivial identity makes the computation of stream derivatives
often surprisingly simple.

\medskip
\noindent

\begin{example}
\label{examples of rational streams}
Let
\[
\sigma =
\frac{1}{1- (c \times X)}
\]
with $c \in A$. We compute
\begin{eqnarray*}
\sigma '
& = &
\left( \sigma - \sigma(0) \right) '
\comment{Corollary \ref{calculating derivatives}}
\\
& = &
\left( \frac{1}{1- (c \times X)} - 1  \right) '
\comment{Proposition  \ref{basic properties of stream calculus}}
\\
& = &
\left( \frac{c \times X}{1- (c \times X)}  \right) '
\\
& = &
\left( \, X \times  \frac{c}{1- (c \times X)}  \right) '
\comment{identity (\ref{multiplying with X})}
\\
& = &
\frac{c }{1- (c \times X)}
\comment{identity (\ref{derivative of X times sigma})}
\end{eqnarray*}
and, more generally,
\begin{eqnarray*}
\sigma^{(n)}
& = &
\frac{c^n }{1- (c \times X)}
\end{eqnarray*}
Using the fact that $\sigma(n) = \sigma^{(n)} (0)$,
this yields the following well-known expression
for this
prototypical example of a rational stream:
\begin{equation}
\label{the prototypical rational stream}
\frac{1}{1- (c \times X)} = (1, \,c, \,c^2, \, \ldots )
\end{equation}
Similarly for
\[
\tau =
\frac{1}{(1- X)^2}
\]
one computes
\begin{eqnarray*}
\tau '
& = &
\left( \tau - \tau(0) \right) '
\\
& = &
\left( \frac{1}{(1- X)^2} - 1  \right) '
\\
& = &
\left( \frac{2X - X^2}{(1- X)^2} \right) '
\\
& = &
\left( \, X \times  \frac{2 - X}{(1- X)^2} \right) '
\\
& = &
\frac{2 - X}{(1- X)^2}
\end{eqnarray*}
and again more generally,
\[
\tau^{(n)} =
\frac{(n+1) - (n\times X)}{(1- X)^2}
\]
leading to
\[
\frac{1}{(1- X)^2}
= \,
(1,\, 2,\, 3,\, \ldots )
\]
\qed
\end{example}

\section{Streams and vector spaces}
\label{Streams and vector spaces}

Let $V$ be a vector space (over a field $k$).
The set
$$
V^\omega = \,
\{ \sigma \mid \sigma : \{0,1,2, \ldots \} \to V \}
$$
of streams over $V$
is itself a vector space, with addition and
scalar multiplication given, for $n \geq 0$ and $c \in k$, by
$$
(\sigma + \tau )(n) = \sigma (n) +  \tau (n)
\;\;\;\;\;\;\;
(c \cdot \sigma )(n) = c \cdot \sigma (n)
$$
where on the right we use vector addition and scalar multiplication
in the vector space $V$.

For future reference, we denote
the operations of initial value and
derivative by
$i: V^\omega \to V$ and $d: V^\omega \to V^\omega$:
for all $\sigma \in V^\omega$,
\[
i(\sigma) = \sigma(0)
\;\;\;\;\;\;\;
d(\sigma) = \sigma '
\]

\begin{proposition}
\label{head and tail are linear transformations}
{\rm
The operations of initial value
$i: V^\omega \to V$ and derivative
$d: V^\omega \to V^\omega$ are linear.

\begin{proof}
Immediate from
$$
(x \cdot \sigma + y \cdot \tau)(0) =
x \cdot \sigma(0)  + y \cdot \tau(0)  \;\;\;\;\;
(x \cdot \sigma + y \cdot \tau)' =
x \cdot \sigma' + y \cdot \tau'
$$
for all $x,y \in k$ and
$\sigma,\tau \in V^\omega$.
\end{proof}
}
\end{proposition}

\subsection{Streams of linear transformations}
Next we define streams of linear transformations.
To this end, we first note that the set
\[
L(V,V) = \{ F: V \to V \mid  \,
\mbox{$F$ is a linear transformation} \; \}
\]
(which we shall usually denote by $L$)
is a ring
\[
(L,+_L, \, \cdot_L , \, 0_L , \, 1_L \, )
\]
Addition $F+_LG$, multiplication
$F \cdot_L G$, and negation $-_L F$ are defined, for all $F,G \in L$ and
$v \in V$, by
\begin{eqnarray*}
(F+_LG)(v)
& = &
 F(v) + G(v)
\\
(F \cdot_L G )(v)
& = &
F \circ G  (v)
\\
(-_L F )(v)
& = &
-F(v)
\end{eqnarray*}
The
neutral elements $0_L: V \to V$ and $1_L:V \to V$
for sum and multiplication are given, for all $v \in V$, by
$0_L(v) = 0_V$ (the zero vector in $V$) and
$1_L (v) = v$.

The set $L$ of linear transformations on
a vector space $V$ being a ring,
we have, by the definitions of
Section \ref{Rational streams},
a calculus of streams of linear transformations.
Streams $\phi \in L^\omega$
are infinite sequences $\phi = (\phi(0), \phi(1), \phi(2) , \ldots )$
of linear transformations $\phi(i): V \to V$.
The operations of sum and (convolution) product
are given, for $\phi,\psi \in L^\omega$, by
\begin{eqnarray*}
(\phi + \psi)(n) & = &
\phi (n) +_L \psi(n)
\\
(\phi \times \psi)(n) & = &
\sum_{i=0}^{n}
\phi (i) \cdot_L \psi(n-i)
\\
& = &
\sum_{i=0}^{n}
\phi (i) \circ \psi(n-i)
\end{eqnarray*}
As before we also have,
for every $F \in L$,
a constant stream
\[
[F] = ( F , \, 0_L, \, 0_L , \, 0_L , \, \ldots)
\]
In particular we also have
\[
[1_L] = ( 1_L , \, 0_L, \, 0_L , \, 0_L , \, \ldots)
\]
which we shall often simply denote by $1$.
The constant stream $X$ now looks like
\[
X = (0_L,\, 1_L , 0_L , \, 0_L , \, 0_L , \, \ldots )
\]
Every stream $\phi \in L^\omega$ has an additive inverse
$-\phi$ given, as before, by
\[
-\phi =
(- \phi(0), \, - \phi(1), \, - \phi(2) , \, \ldots )
\]
A stream $\phi \in L^\omega$ has a
(unique) multiplicative inverse
$\phi^{-1}$ in $L^\omega$:
\[
\phi^{-1} \times \phi = 1 \;\; (= \, [1_L] \, )
\]
whenever the linear transformation
$\phi(0) : V \to V$ has a multiplicative
inverse in the ring $L$, that is, whenever
$\phi(0)$ is invertible.

\begin{example}
\label{example of rational stream of transformations}
For any linear transformation $F \in L$,
we define the stream $\tilde{F} \in L^\omega$ by
\[
\tilde{F} = \;
\frac{1}{1 - ([F] \times X)}
\]
As before, it is a prototypical example of a rational stream.
Note that
\[
1 - ([F] \times X) = (1_L , \, -F , \, 0_L , \, 0_L , \, 0_L , \, \ldots)
\]
indeed {\em is} invertible in $L^\omega$ as $1_L$ is trivially
invertible in $L$. As an instance of identity
(\ref{the prototypical rational stream})
in Example \ref{examples of rational streams}, we have
\begin{equation}
\label{F tilde is rational}
\tilde{F}
= \;
\frac{1}{1 - ([F] \times X)}
= \;
(1, F , F^2, \cdots )
\end{equation}
where now $F^{n+1} = F \circ F^n$, all $n \geq 0$.
\qed
\end{example}

\medskip
So far we have looked at the set $L=L(V,V)$
of streams of linear transformations
from a vector spave $V$ to itself. It will also be convenient
to consider (streams of)  linear transformations between
two different vector spaces $V$ and $W$:\[
L(V,W) = \{ F: V \to W \mid  \,
\mbox{$F$ is a linear transformation} \; \}
\]
Note that $L(V,W)$ is not a (semi-)ring --- we cannot
define multiplication to be composition as we did with
$L(V,V)$ --- and as a consequence the set $L(V,W)^\omega$ of streams
over $L(V,W)$ does not have as much structure
as the set $L(V,V)^\omega$. It will be convenient, however,
to use the following generalised version of the
operation of convolution product. For
vector spaces $U,V,W$ and
for all $\phi \in L(U,V)^\omega$ and $\psi \in L(V,W)^\omega$
we define $\psi \times \phi \in L(U,W)^\omega$, for all $n \geq 0$, by
\begin{equation}
\label{generalised convolution product}
(\psi \times \phi)(n) = \,
\sum_{i=0}^n \psi(i) \circ \phi(n-i)
\end{equation}
One can also apply streams of linear transformations to
streams of vectors, as follows.
For all $\phi \in L(V,W)^\omega$ and $\sigma \in V^\omega$
we define $\phi \times \sigma \in W^\omega$, for all $n \geq 0$,
by
\begin{equation}
\label{convolution of functions and streams}
(\phi \times \sigma)(n) = \,
\sum_{i=0}^n \phi(i) \left(  \sigma(n-i) \right)
\end{equation}
For a linear transformation $H: V \to W$, we put again
\[
[H] = \, (H,\,  0_L,\, 0_L,\, 0_L,\,  \ldots)
\]
(where now $0_L$ is the everywhere zero transformation
from $V$ to $W$).
As a special case of (\ref{convolution of functions and streams})
we have
\[
[H] \times \sigma =
(H(\sigma(0)), \, H(\sigma(1)), \,
H(\sigma(2)), \ldots)
\]
Note that the set of streams
$L(V,V)^\omega$ has also its own operation of
convolution product, which interacts nicely
with the product defined in
(\ref{convolution of functions and streams}).
For instance, for $\phi, \psi \in L(V,V)^\omega$
and $\sigma \in V^\omega$,
\begin{equation}
\label{convolution product and convolution product}
(\phi \times \psi ) \times \sigma =
\phi  \times ( \psi  \times \sigma )
\end{equation}


\subsection{Streams of matrices}
Since linear transformations between finite dimensional
vector spaces (over a field $k$) correspond to
matrices (with entries in $k$), streams of linear transformations
correspond to streams of matrices. Here we show how
{\em rational} streams of linear transformations
correspond to matrices with
rational streams (over $k$) as entries.

First some conventions.
For any set $A$ and $n \geq 1$, we denote the elements  $v \in A^n$
by $ v=(v_1, \ldots , v_n) $.
It will sometimes be convenient to switch between
streams of tuples and tuples of streams.
We define the {\em transpose\/} as follows:
\begin{equation}
\label{definition streams to vectors}
(-)^T: (A^n)^\omega \to (A^\omega)^n \;\;\;\;\;\;\;
(\sigma^T)_i(j) = (\sigma(j))_i
\end{equation}
This function is an isomorphism and has an inverse
which we denote again by
$$
(-)^{T} : (A^\omega)^n \to (A^n)^\omega
$$

Now let $k$ be a field.
A linear transformation $F: k^n \to k^m$
between {\em finite dimensional\/} vector spaces
corresponds to an $m \times n$ matrix $M_F$ with values
$F_{ij}$ in $k$:
$$
F: k^n \to k^m
\;\;\;\;\;\;\;\;\;\;\;\;
M_F =
\left(
\begin{array}{cccc}
F_{11} &
F_{12} &
\cdots &
F_{1n}
\\
F_{21} &
F_{22} &
\cdots &
F_{2n}
\\
\vdots &
\vdots &
\ddots &
\vdots
\\
F_{m1} &
F_{m2} &
\cdots &
F_{mn}
\end{array}
\right)
$$
Here and in what follows, the matrix is with respect
to the standard basis
\[
(1,0, \ldots , 0), \; \ldots , \; (0,\ldots , 0, 1)
\]
of $k^n$ and $k^m$.
Any stream
$\phi = (\phi(0),\, \phi(1),\, \phi(2) , \, \ldots)$
of linear transformations $\phi(i): k^n \to k^m$
corresponds to a stream of matrices
\[
(M_{\phi(0)},\, M_{\phi(1)},\, M_{\phi(2)} , \, \ldots)
=
M_{\phi(0)} \,  + \, (M_{\phi(1)} \times X)
\, + \,  (M_{\phi(2)}  \times X^2)  \, +  \cdots
\]
If we consider $M_{\phi(i)}  \times X^i$ as an
$m \times n$ matrix obtained from $M_{\phi(i)}$ by
multiplying each of its entries by $X^i$, then the infinite
sum on the right can itself be viewed as an $m \times n$
matrix $M_\phi$ with entries in $k^\omega$:
\begin{equation}
\label{definition M phi}
(M_\phi)_{ij}
= \;
(M_{\phi(0)})_{ij}  \,  + \, ((M_{\phi(1)})_{ij}  \times X)
\, + \,  ((M_{\phi(2)})_{ij}   \times X^2)  \, +  \cdots
\end{equation}
The correspondence between $\phi$ and $M_\phi$ is given by the
following commutative diagram:
\begin{equation}
\label{streams to vectors}
\xymatrix{
(k^n)^\omega \rto^{\phi \times (-)} \dto_-{(-)^T}
& (k^m)^\omega \dto^-{(-)^T}
\\
(k^\omega)^n \rto_-{M_\phi \times (-)} &
(k^\omega)^m }
\;\;\;\;\;\;\;
(\phi \times \sigma)^T =
M_\phi \times \sigma^T
\end{equation}
Here $\phi \times \sigma$ is as defined in
(\ref{convolution of functions and streams}) and
$M_\phi \times \sigma^T$ denotes
matrix to vector multiplication.

Consider $L =  L(k^n,k^n)$ and recall that
$[1_L] = (1_L, \, 0_L, \, 0_L, \, 0_L, \, \ldots )$.
Let $I$ be the $n \times n$ identity matrix over $k$. We have:
\begin{equation}
\label{unit law for streams and matrices}
M_{[1_L]} = I
\end{equation}
Furthermore addition and convolution product
of streams of linear transformations, on the one hand,
and matrix addition and multiplication, on the other, are related as follows:
\[
M_{\phi + \psi} = M_\phi + M_\psi
\]
\begin{equation}
\label{convolution product corresponds to matrix multiplication}
M_{\phi \times \psi} = M_\phi \times M_\psi
\end{equation}
As a consequence, we have the following proposition.

\begin{proposition}
\label{matrix of rational stream has rational entries}
Let $\rho \in L(k^n, k^n)^\omega$
be a stream of linear transformations
$\rho(i): k^n \to k^n$.
If $\rho$ is rational
then $M_\rho$ defined in (\ref{definition M phi})
has entries in $Rat(k^\omega)$.
\end{proposition}

\bigskip
\noindent
{\bf Proof:\/}
Consider two polynomial streams
$\phi, \psi \in L(k^n,k^n)^\omega$.
The entries of the matrices $M_\phi$ and
$M_\psi$ are polynomial streams
in $k^\omega$. If $\psi$ moreover
has an inverse $\psi^{-1}$ then
(\ref{unit law for streams and matrices}) and
(\ref{convolution product corresponds to matrix multiplication})
imply $M_{\psi^{-1}} = (M_\psi)^{-1}$, which
has values in $Rat(k^\omega)$.
It follows that
$M_{\phi \times \psi^{-1}} =
M_\phi \times (M_\psi)^{-1}$ has values in $Rat(k^\omega)$.
\qed

\bigskip
\noindent

\begin{example}
\label{example linear system of differential equations}
{\rm
Let $k = \ofR$ and let
$F,G: \ofR^2 \to \ofR^2$ be linear transformations
defined by
$$
M_F = \,
\left(
\begin{array}{cc}
1 & 1
\\
0 & 0
\\
\end{array}
\right)
\;\;\;\;\;\;\;\;
M_G = \,
\left(
\begin{array}{cc}
0 & -1
\\
1 & 2
\\
\end{array}
\right)
$$
We compute the matrices of the rational streams
$\tilde{F} = (\, 1 - (F \times X) \,)^{-1}$  and
$\tilde{G} = (\, 1 - (G \times X) \,)^{-1}$:
$$
M_{\tilde{F}} = \;
(M_{1 - (F \times X)})^{-1} = \;
\left(
\begin{array}{cc}
1-X \; \; & -X
\\
0 \;  \;& 1
\\
\end{array}
\right)^{-1}
\; = \;
\left(
\begin{array}{cc}
\frac{1}{1-X} \; \; & \frac{X}{1-X}
\\
0 \; \; & 1
\\
\end{array}
\right)
$$
$$
M_{\tilde{G}} = \;
(M_{1 - (G \times X)})^{-1} = \;
\left(
\begin{array}{cc}
1 \; \; & X
\\
-X \; \; & 1-2X
\\
\end{array}
\right)^{-1}
\; = \;
\frac{1}{(1-X)^2} \, \times \, \left(
\begin{array}{cc}
1-2X \; \; & -X
\\
X \; \; & 1
\\
\end{array}
\right)
$$
\qed
}
\end{example}

\section{Linear representations}
\label{Linear representations}

We introduce linear systems and show
how they can be used as representations for streams.
In particular, we shall show how {\em finite dimensional}
linear systems represent {\em rational} streams.

Let $O$ be a vector space the elements of which
we think of as {\em outputs}.
A {\em linear system with output in $O$}
is a pair $(V,<H,F>)$
consisting of a vector space $V$ called the {\em state space\/}
together with a linear transformation
$F: V \to V$ called the {\em transition function} (or dynamics)
and a linear transformation
$H: V \to O$ called the {\em output function}.

A linear system with output in $O$ --- or linear
$O$-system for short --- is in other words
a {\em coalgebra} of the functor
\[
O \times (-) \, : Vect \to Vect
\]
on the category $Vect$ of vector spaces and linear transformations.
As a consequence, there is the following (standard) notion
of homomorphism.
A {\em homomorphism of linear systems}
$(V,<H_V,F_V>)$ and $(W,<H_W,F_W>)$
is a linear transformation $f: V \to W$ such that
$H_W \circ f = H_V$ and $F_W \circ f = f \circ F_V$:
$$
\xymatrix{
V \ar@{->}[r]^f   \dto_{<H_V,F_V>}  & W \dto^{<H_W,F_W>}
\\
O \times V \rto^{1 \times f}    & O \times W
}
$$

We saw (in Section \ref{Streams and vector spaces}) that if
$O$ is a vector space then $O^\omega$ is also a vector
space. Since the operations of initial value
$i: O^\omega \to O$ and derivative
$d: O^\omega \to O^\omega$ are
linear transformations
(Proposition \ref{head and tail are linear transformations}),
$(O^\omega, <d,i>)$ is a linear $O$-system.
It is {\em final} among all linear $O$-systems.

\begin{proposition}[Finality]
\label{finality}
{\rm
From every linear $O$-system
$(V,<H,F>)$ there exists precisely one homomorphism
to $(O^\omega, <i,d>)$:
$$
\xymatrix{
V \ar@{-->}[r]^f   \dto_{<H,F>}  & O^\omega \dto^{<i,d>}
\\
O \times V \rto^{1 \times f}    & O \times O^\omega
}
$$
}
\end{proposition}

\begin{proof}
There exists precisely one {\em function} $f:V \to O^\omega$
making the diagram above commute. It is given by
\[
f(v) = \, (H(v), \, H \circ F(v), \, H \circ F^2(v) , \ldots )
\]
for all $v \in V$ and it is linear because both $H$ and $F$ are.
\end{proof}

\medskip
\begin{definition}[Linear representation]
\label{Linear representation}
{\rm
In the situation above, we call the stream
$f(v)$ the {\em final behaviour\/}
of $v$.
We call the linear $O$-system $(V,<H,F>)$
with designated point $v \in V$
a {\em linear representation} for the stream $\sigma \in O^\omega$
if $f(v) = \sigma$.
\qed

}

\end{definition}

\medskip

Next we look at the special case where both $O$ and $V$ are
{\em finite dimensional} vector spaces over $k$.
So let $n,m \geq 1$, let
\[
O = k^m, \;\;\;\; V = k^n
\]
and consider a linear $k^m$-system $(k^n , <H,F>)$ with dynamics
$F: k^n \to k^n$ and output $H: k^n \to k^m$.
The final behaviour $f:k^n \to (k^m)^\omega$
will map any state $v \in k^n$ to a stream of vectors in $k^m$.
We claim that the transpose of the latter consists of a
vector of $m$ rational streams in $k^\omega$.

\begin{theorem}
\label{characterising finite dimensional linear O-systems}
{\rm
Let $n,m \geq 1$ and let $(k^n,<H,F>)$
be a finite dimensional $k^m$-system.
Let
\[
f: k^n \to (k^m)^\omega
\]
be the final behaviour homomorphism.
Then for all $v \in k^n$,
\[
f(v)^T \in Rat(k^\omega)^m
\]
}
\end{theorem}

\begin{proof}
First we observe that for every $v \in V$,
we can express $f(v)$ in terms of convolution products
as follows:
\begin{eqnarray*}
f(v) & = &
(H(v), \, H \circ F(v), \, H \circ F^2(v) , \ldots )
\\
& = &
(H, 0,0,0, \ldots ) \times (1,F,F^2 , \ldots ) \times
(v , 0,0,0, \ldots )
\comment{using
(\ref{convolution of functions and streams})
and (\ref{convolution product and convolution product})}
\\
& = &
(H, 0,0,0, \ldots ) \times \tilde{F} \times
(v , 0,0,0, \ldots )
\comment{using (\ref{F tilde is rational})}
\\
& = &
[H] \times \tilde{F} \times [v]
\end{eqnarray*}
By (\ref{streams to vectors}),
the following diagram
commutes:
\begin{equation}
\xymatrix@C+3.0ex{
(k^n)^\omega \rto^{\tilde{F} \times (-)} \dto_-{(-)^T}
& (k^n)^\omega \dto^-{(-)^T}
\rto^{[H] \times (-)}
& (k^m)^\omega \dto^-{(-)^T}
\\
(k^\omega)^n \rto_-{M_{\tilde{F}} \times (-)} &
(k^\omega)^n
\rto_-{M_{[H]} \times (-)} &
(k^\omega)^m }
\end{equation}
or, equivalently,
\[
([H] \times \tilde{F} \times (-))^T =
M_{[H]} \times M_{\tilde{F}} \times (-)^T
\]
It follows that the final behaviour $f$
satisfies
\begin{equation}
\label{final behaviour is rational}
f(v)^T \, = \,
( [H] \times \tilde{F} \times [v] )^T
\, = \,
M_{[H]} \times M_{\tilde{F}} \times [v]^T
\end{equation}
The matrix $M_{[H]}$ has entries in ($k$ and thus in)
$Rat(k^\omega)$.
Since $\tilde{F} = (1-(F \times X))^{-1}$ is a rational stream,
the matrix $M_{\tilde{F}}$ has
values in $Rat(k^\omega)$,
by Proposition
\ref{matrix of rational stream has rational entries}.
As a consequence, $f(v)^T$ is obtained from $[v]^T$
by multiplication with an $m \times n$ matrix with
values in $Rat(k^\omega)$. This proves the theorem.
\end{proof}

\medskip
\noindent
Since finite dimensional linear systems are
finitary objects (being completely determined by
two finite matrices),
the relevance of
Theorem \ref{characterising finite dimensional linear O-systems}
lies in the fact that it shows that such finitary systems
represent (vectors of) rational streams.
In Section \ref{Constructing linear representations},
we will see that any rational stream
can be represented in this manner. But first we look at a few
examples illustrating the present theorem.

\begin{example}
\label{example linear O-system}
{\rm
Let $k = \ofR$ and consider the linear system
$(\ofR^2, <H,F>)$ with output $H: \ofR^2 \to \ofR$ and
dynamics $F : \ofR^2 \to \ofR^2$ given by
$$
H =
\left(
\begin{array}{cc}
1 & 1
\end{array}
\right)
\;\;\;\;\;\;\;
F =
\left(
\begin{array}{cc}
1  & 1
\\
0 & 0
\end{array}
\right)
$$
The matrix $M_{\tilde{F}}$ corresponding to
$\tilde{F}$
has been computed
in Example \ref{example linear system of differential equations}:
$$
M_{\tilde{F}} = \;
\left(
\begin{array}{cc}
\frac{1}{1-X}  & \frac{X}{1-X}
\\
0 & 1
\\
\end{array}
\right)
$$
The final behaviour
$f_{<H,F>}: \ofR^2 \to \ofR^\omega$
of this system is given, for any $(a,b) \in \ofR^2$, by
\begin{eqnarray*}
f_{<H,F>}(a,b) & = &
\left(
\begin{array}{cc}
1 & 1
\end{array}
\right)
\, \times \,
\left(
\begin{array}{cc}
\frac{1}{1-X}  & \frac{X}{1-X}
\\
0 & 1
\end{array}
\right)
\, \times \,
\left(
\begin{array}{c}
a
\\
b
\end{array}
\right)
\\
& = &
\frac{a+b}{1-X}
\end{eqnarray*}
(omitting square brackets around $a$ and $b$ as usual).
Repeating the example with a different output function $\bar{H}$
and the same dynamics $F$:
$$
\bar{H} =
\left(
\begin{array}{cc}
1 & 2
\end{array}
\right)
\;\;\;\;\;\;\;
F =
\left(
\begin{array}{cc}
1  & 1
\\
0 & 0
\end{array}
\right)
$$
leads to the following final behaviour:
$$
f_{<\bar{H},F>} (a,b)
\;  = \;
\left(
\begin{array}{cc}
\frac{1}{1-X}  & \frac{2-X}{1-X}
\end{array}
\right)
\, \times \,
\left(
\begin{array}{c}
a
\\
b
\end{array}
\right)
\\
\;  = \;
\frac{a+2b -bX}{1-X}
$$
\qed
}
\end{example}

\medskip
Because  linear $O$-systems are coalgebras, the
general definition of coalgebraic equivalence applies.
In conclusion of this section,
we spell out this definition together with
the observation that the corresponding minimization
of a system is given by (the image under) the final
behaviour mapping.

Equivalence of linear
$O$-systems is defined as follows.
A relation $R \subseteq V \times W$ is called
an {\em $O$-bisimulation\/} between $O$-systems
$(V,<H_V,F_V>)$ and $(W,<H_W,F_W>)$ if for all
$v \in V$ and $w \in W$:
$$
<v,w> \in R \Rightarrow
\left \{
\begin{array}{c}
H_V(v) = H_W(w)
\;\;\mbox{and}
\\
<F_V(v),F_W(w)> \in R
\end{array}
\right .
$$
We say that $v$ and $w$ are {\em $O$-equivalent\/} and write
$v \sim_O w$ if there exists an $O$-bisimulation $R$
with $<v,w> \in R$.
The final behaviour $f: V \to O^\omega$ of
an $O$-system $(V,<H_V,F_V>)$ identifies precisely all $O$-equivalent
states:
$v_1 \sim_O v_2$ iff
$f(v_1) = f(v_2)$, for all $v_1,v_2 \in V$.
(For the elementary proof, see \cite{Rut05b}.)
As a consequence, the minimization of an $O$-system with respect
to $O$-equivalence is given by
the image of $V$ under $f$, which is a subsystem
$f(V) \subseteq O^\omega$ because $f$ is a homomorphism.
It follows that the greatest $O$-equivalence
on $V$ is given by the kernel $ker(f)$.

\section{Constructing linear representations}
\label{Constructing linear representations}

Let $k$ be a field.
We show how to construct finite-dimensional
linear representations for
(vectors of) rational streams in $k^\omega$.

For a stream $\sigma \in O^\omega$ we consider
the smallest
{\em subspace\/} of $O^\omega$ that contains $\sigma$ and is closed
under the operation of stream derivative,
that is, the linear transformation $d: O^\omega \to O^\omega$.
This (so-called $d$-cyclic) vector space $Z_\sigma$
is the subspace of $O^\omega$
that is spanned by the set of vectors given by
\begin{equation}
\label{states generated by sigma}
\{ \sigma^{(0)} , \, \sigma^{(1)} , \, \sigma^{(2)} , \, \ldots \}
\end{equation}
with $\sigma^{(0)} = \sigma$ and $\sigma^{(n+1)} =
d(\sigma^{(n)}) =  (\sigma^{(n)})'$.
We can turn $Z_\sigma$ into a linear system by taking
as output function and transition function
the restrictions of
$i: O^\omega \to O$ and $d:  O^\omega \to O^\omega$ to $Z_\sigma$.
The set inclusion
\[
f: \, Z_\sigma \subseteq O^\omega
\]
is then a homomorphism
of linear $O$-systems. By finality of
$(O^\omega, <i,d>)$, this homomorphism
is unique. It follows that
$(Z_\sigma,<i,d>)$ with initial state $\sigma$
is a minimal representation of $\sigma$.

In general, the dimension of $Z_\sigma$ will be infinite.
Of special interest are those $\sigma \in O^\omega$ for which
there exists an $n \geq 1$ such that
all of $ \sigma = \sigma^{(0)}$ through $\sigma^{(n-1)}$
are linearly independent and
$$
\sigma^{(n)} \, = \;
\sum_{i=0}^{n} \, c_i \times \sigma^{(i)}
$$
for some coefficients $c_0 , \ldots , c_{n-1}$
in the base field $k$ of $O$ and $O^\omega$.
Then $Z_\sigma$ is a vector space of dimension $n$. The
linear transformation $G: Z_\sigma \to Z_\sigma$ induced by
$d: O^\omega \to O^\omega$ is given, with respect to the
(ordered) basis $\sigma^{(0)}, \ldots , \sigma^{(n-1)}$, by
the $n \times n$ matrix
$$
M_G = \;
\left(
\begin{array}{ccccc}
0 \;  & 0 \;  & \cdots & 0 \;  & c_0
\\
1 \;  & 0 \;  & \cdots & 0 \;  & c_1
\\
0 \;  & 1 \;  & \cdots & 0 \;  & c_2
\\
\vdots &
\vdots &
\ddots &
\vdots &
\vdots
\\
0 \;  & 0 \;  & \cdots & 1 \;  & c_{n-1}
\end{array}
\right)
$$
(This matrix is in fact (a variation of)
the {\em companion\/} matrix of the
so-called $d$-order polynomial of $\sigma$; cf.
\cite[Thm.15, p.339]{BML77}.)
The linear transformation $H: Z_\sigma \to O$ induced by
$i: O^\omega \to O$ is given, again with respect to the
basis $\sigma^{(0)}, \ldots , \sigma^{(n-1)}$, by
the matrix (of size $dim(O) \times n$)
$$
M_H = \;
\left(
\begin{array}{ccccc}
\sigma^{(0)}(0) \; \; &
\sigma^{(1)}(0) \; \; &
\sigma^{(2)}(0) \; \; &
\cdots  \;&
\sigma^{(n-1)}(0)
\end{array}
\right )
$$
Thus we have obtained a
linear $O$-system $(Z_\sigma , <H,G>)$ of dimension $n$.
As before, the inclusion $f: Z_\sigma \subseteq O^\omega$
is a homomorphism. Thus
$f(\tau) = \tau$, for all $\tau \in Z_\sigma$ and
$(Z_\sigma , <H,G>)$ with $\sigma$ as initial state is a minimal
representation of $\sigma$.

\bigskip
\noindent
\begin{example}
\label{example constructing linear representations}
{\rm
Let $O = \ofR$ and
consider the stream
$\sigma = 1/(1-X)^2 \in O^\omega$.
Computing the successive stream derivatives
of $\sigma = \sigma^{(0)}$,
using Corollary \ref{calculating derivatives},
gives
$$
\sigma^{(1)}
= \,
\frac{2-X}{(1-X)^2}
\;\;\;\;\;\;\;
\sigma^{(2)}
= \,
\frac{3-2X}{(1-X)^2}
= \,  - \sigma^{(0)} \, + \, (2 \times \sigma^{(1)})
$$
Thus $\sigma^{(0)}$ and $\sigma^{(1)}$ form a basis
for $Z_\sigma$. Because
$\sigma^{(0)} (0) = 1$ and $\sigma^{(1)}(0) = 2$, we have
$$
M_H = \left(
\begin{array}{cc}
1 & 2
\end{array}
\right)
\;\;\;\;\;\;\;\;
M_G = \;
\left(
\begin{array}{cc}
0  & -1
\\
1 & 2
\end{array}
\right)
$$
Now $\sigma$ is represented by $(Z_\sigma , <H,G>)$,
with $\sigma$ as the initial state.
Clearly, $\ofR^2 \cong Z_\sigma$. Note that this
isomorphism can also be obtained by computing
the final behaviour $f: \ofR^2 \to \ofR^\omega$
of the $O$-system $(\ofR^2 , <H,G>)$, using
Theorem \ref{characterising finite dimensional linear O-systems}.
This gives, for all $(r_1,r_2) \in \ofR^2$,
\begin{eqnarray*}
f(r_1,r_2) & = &
M_H \times M_{\tilde{G}} \times (r_1,r_2)
\\
& = &
\left(
\begin{array}{cc}
1\; & \; 2
\end{array}
\right)
 \; \times \;
\frac{1}{(1-X)^2}
 \; \times \,
\left(
\begin{array}{cc}
1-2X \;  & \; -X
\\
X \; &  1
\end{array}
\right)
\; \times \;
\left(
\begin{array}{c}
r_1 \\ r_2
\end{array}
\right)
\end{eqnarray*}
(Recall the computation of $M_{\tilde{G}}$ from
Example \ref{example linear system of differential equations}.)
As expected, we have $f(1,0) = \sigma$
and $f(0,1) = \sigma^{(1)}$.
\qed
}
\end{example}

\bigskip
\noindent
\begin{example}
{\rm
Let $O = \ofR^2$ and
consider the pair
$(\tau , \sigma) \in (\ofR^\omega)^2 \cong (\ofR^2)^\omega$,
with
$\tau = 1/(1-2X)$ and $\sigma = 1/(1-X)^2$.
Computing (pairs of) stream derivatives
\begin{eqnarray*}
(\tau , \sigma)^{(1)}
& = &
\left( \, \frac{2}{1-2X} , \; \frac{2-X}{(1-X)^2} \, \right)
\\
(\tau , \sigma)^{(2)}
& = &
\left( \, \frac{2^2}{1-2X} , \; \frac{3-2X}{(1-X)^2} \, \right)
\\
(\tau , \sigma)^{(3)}
& = &
\left( \, \frac{2^3}{1-2X} , \; \frac{4-3X}{(1-X)^2} \, \right)
\\
& = &
2 \times (\tau , \sigma)^{(0)}
\, - \,  5 \times (\tau , \sigma)^{(1)}
\, + \, 4 \times (\tau , \sigma)^{(2)}
\end{eqnarray*}
we see that $Z_{(\tau, \sigma)}$ has dimension $3$ with
$H: Z_{(\tau, \sigma)} \to \ofR^2$ and
$G : Z_{(\tau, \sigma)} \to Z_{(\tau, \sigma)}$
given by
$$
M_H = \left(
\begin{array}{ccc}
1 & \; 2 \; & 4
\\
1 & 2 & 3
\end{array}
\right)
\;\;\;\;\;\;\;\;
M_G = \;
\left(
\begin{array}{ccc}
0 & 0 & 2
\\
1 & \; 0 \; & \; -5
\\
0 & 1 & 4
\end{array}
\right)
$$
\qed
}
\end{example}

\bigskip
\noindent
\begin{theorem}
\label{realisation of rational streams}
Let $k$ be a field and let $O = k^m$.
A vector of streams $\sigma \in (k^\omega)^m$
is representable by a linear $k^m$-system of finite dimension
iff $\sigma \in (Rat(k^\omega))^m$.
\end{theorem}

\begin{proof}
From left to right, this is
Theorem \ref{characterising finite dimensional linear O-systems}.
For the converse, it is sufficient to observe
that the examples above generalise to arbitrary vectors
of rational streams. This is immediate from the fact that
for a rational stream $\sigma = \rho/\tau$,
the dimension of $Z_\sigma$ in the construction
above is bounded by the
maximum of the degrees of $\rho$ and $\tau$.
\end{proof}

\medskip
\noindent
For single streams, the results
of this section can be summarized as follows.

\begin{theorem}
\label{first equivalence theorem}
{\rm
Let $k$ be a field. For a stream $\sigma \in k^\omega$,
the following are equivalent:
\begin{itemize}
\item[(1)]
The stream $\sigma$ is rational: $\sigma = \rho/\tau$
for polynomial streams $\rho$ and $\tau$ (with $\tau(0) \neq 0$).
\item[(2)]
The stream $\sigma$ is representable by a linear system
of finite dimension.
\item[(3)]
The subsystem $Z_\sigma \subseteq (k^\omega,<i,d>)$
generated by $\sigma$ has finite dimension.
\end{itemize}
\qed
}
\end{theorem}

\medskip
\noindent
In conclusion of this section,
we show that (3) above can be conveniently used
to prove that a stream is {\em not} rational.

\begin{corollary}
\label{non rationality}
{\rm
In order to prove that a stream $\sigma \in k^\omega$ is
{\em not} rational, it suffices to show that
\[
\{ \sigma^{(0)} , \, \sigma^{(1)} , \, \sigma^{(2)} , \, \ldots \}
\, \subseteq k^\omega
\]
contains infinitely many linearly independent vectors.
\qed
}
\end{corollary}

\medskip
\begin{example}
\label{example of a non-rational stream}
{\rm
Consider $\sigma \in \ofR^\omega$ given by
\begin{eqnarray*}
\sigma
& = &
(1,\, 1,\, 0,\, 1,\, 0,\, 0,\, 1,\, 0,\, 0,\,  0,\,
1,\, 0,\, 0,\,  0,\,  0,\, \ldots )
\\
& = &
1 + \, X + \, X^3 + \, X^6 + \, X^{10} + \, X^{15} + \, \cdots
\\
& = &
\sum_{k=0}^{\infty}   X^{k(k+1)/2}
\end{eqnarray*}
The set of stream derivatives of $\sigma$ contains
the following infinite subset of linearly independent streams:
\[
(1 , \; \ldots )
\]
\[
(0, \; 1 , \; \ldots )
\]
\[
(0, \; 0, \; 1 , \; \ldots )
\]
\[
(0, \; 0, \; 0, \; 1 , \; \ldots )
\]
\[
\cdots
\]
Thus $\sigma$ is not rational.
}
\qed
\end{example}

\section{Stream circuits}
\label{Stream circuits}

We saw that rational streams can be represented by
finite dimensional linear systems.
Such systems are finitary in that they are determined
by (two) finite dimensional matrices with values in $k$.
In this section, we show that such systems
--- and as a consequence rational streams --- are,
equivalently, computed by so-called stream circuits with finite
memory.

{\em Stream circuits} (with values in a field $k$)
are data flow networks that act on streams of inputs and
produce streams of outputs. They are built out of four
types of basic gates by means of composition, which amounts
simply to connecting (single) output ends to (single) input ends.
We introduce these basic gates below, first describing
their single-step behaviour in terms of input and output
values (in $k$). Next we shall describe their
behaviour in terms of input and output streams (in $k^\omega$).
\begin{itemize}

\item[(i)]
For a fixed $r \in k$, an {\em ${r}$-multiplier}
$$
\xymatrix@+4.0ex{ x \ar@{|->}[r]^{{r}} & {r} \cdot x }
$$
inputs a value $x \in k$ at its input end
and outputs that value multiplied with $r$ at its output end.

\item[(ii)]
A {\em register}
$$
\xymatrix@C+6.0ex{
y
\ar @{|->}[r]|(.5)*+[F]{{x}}
&
}
$$
is a one-element buffer (or memory cell) containing
as initial value an element $x \in k$.
Its stepwise computation consists of the (simultaneous)
output of the present value $x$ in the buffer together with the input
of an element $y \in k$, which becomes the new contents of the buffer:
$$
\xymatrix@C+6.0ex{
\ar @{|->}[r]|(.5)*+[F]{{y}}
&
x
}
$$

\item[(iii)]
An {\em adder}
$$
\xymatrix@-4.0ex@C+3.0ex{
x \ar @/^0.5pc/ @{|-}[dr] & &
\\
& {+} \ar@{->}[r] & x {+} y
\\
y \ar @/_0.5pc/ @{|-}[ur] &
}
$$
takes two input values at its input ends and
outputs their sum at its output end.
Here we show
a $2$-to-$1$ adder but more generally we will
also use $n$-to-$1$ adders, for $n \geq 2$.

\item[(iv)]
Lastly a {\em copier}
$$
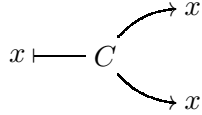
\xymatrix@-4.0ex@C+3.0ex{ & & x \\
x \ar@{|-}[r] & {C} \ar @/^0.5pc/ @{->}[ur]
\ar @/_0.5pc/ @{->}[dr] & \\ & & x }
$$
inputs a value at its input end and outputs multiple
copies of it at its output ends.
Here we show
a $1$-to-$2$ copier but more generally we will
also use $1$-to-$n$ copiers, for $n \geq 2$.

\end{itemize}
Sometimes it will be convenient to combine
multipliers with adders (and similarly copiers).
For instance,
$$
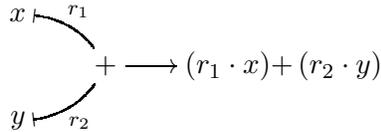
\xymatrix@-4.0ex@C+3.0ex{
x \ar @/^0.5pc/ @{|-}[dr]^-{r_1} & &
\\
& {+} \ar@{->}[r] & (r_1 \cdot x) {+} \,  (r_2 \cdot y)
\\
y \ar @/_0.5pc/ @{|-}[ur]_-{r_2} &
}
$$
multiplies its inputs $x$ and $y$ with the values
$r_1$ and $r_2$ and outputs the sum of the results.

The presence of memory (in the form of registers)
makes that the behaviour of stream circuits
cannot be described simply in terms of
functions of single input and output values in $k$.
(This is reflected in our explanations above
by the fact that we needed {\em two} pictures to illustrate
the behaviour of a register.)
Rather we shall describe the behaviour of our circuits
in terms of {\em streams\/} of inputs and outputs.
As it turns out, all we need are the basic operations
of stream calculus:

\begin{itemize}

\item[(i)]
An $r$-multiplier converts a stream of inputs $\sigma \in k^\omega$
$$
\xymatrix@+4.0ex{ \blue{\sigma}
\ar@{|->}[r]^-{\red{r}} & \red{[r]} \blue{\times \sigma } }
$$
into a stream of outputs $\red{[r]} \blue{\times \sigma }$
by elementwise multiplying the
input values with $r$:
\[
\red{[r]} \blue{\times \sigma }
= \; ( r \cdot \sigma(0),\, r \cdot \sigma(1) , \, r \cdot \sigma(2) ,
\, \ldots \, )
\]

\item[(ii)]
A register with initial value $x \in k$ takes a stream
of inputs $\sigma$
\begin{equation}
\label{register law}
\xymatrix@C+6.0ex{
\blue{\sigma}
\ar @{|->}[r]|(.3)*+[F]{\red{x}}
&
\red{[x]} \blue{ + \, (X \times \sigma) }
}
\end{equation}
and outputs it with one step delay, after having output
the initial value $x$ first:
\[
\red{[x]} \blue{ + \, (X \times \sigma) }
= \; (\red{x} \blue{,\, \sigma(0),\, \sigma(1) , \, \ldots \, )}
\]

\item[(iii)]

An adder takes two input streams $\sigma$ and $\tau$
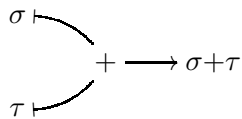
$$
\xymatrix@-4.0ex@C+3.0ex{
\blue{\sigma} \ar @/^0.5pc/ @{|-}[dr] & &
\\
& \red{+} \ar@{->}[r] & \blue{\sigma} \red{+} \blue{\tau}
\\
\blue{\tau} \ar @/_0.5pc/ @{|-}[ur] &
}
$$
and outputs the stream consisting of their elementwise
addition:
\[
\blue{\sigma} \red{+} \blue{\tau} =
(\sigma(0)+\tau(0) , \,
\sigma(1)+\tau(1) , \,
\sigma(2)+\tau(2) , \, \ldots )
\]

\item[(iv)]
The copier simply copies input streams into output streams:
$$
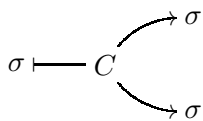
\xymatrix@-4.0ex@C+3.0ex{ & & \blue{\sigma} \\
\blue{\sigma} \ar@{|-}[r] & \red{C} \ar @/^0.5pc/ @{->}[ur]
\ar @/_0.5pc/ @{->}[dr] & \\ & & \blue{\sigma} }
$$

\end{itemize}
Combinations of multipliers and adders (and similarly copiers)
have the expected stream behaviour:
$$
\xymatrix@-4.0ex@C+3.0ex{
\sigma \ar @/^0.5pc/ @{|-}[dr]^-{r_1} & &
\\
& {+} \ar@{->}[r] & ([r_1] \times \sigma) {+} \,  ([r_2] \times \tau)
\\
\tau \ar @/_0.5pc/ @{|-}[ur]_-{r_2} &
}
$$

Now that we have seen the basic gates and their behaviour,
let us look at composite stream circuits and see how their
behaviour can be computed from that of the gates from which
they are made. Consider the following circuit,
built out of two registers, two copiers, three adders, and
six multipliers (two of which
are combined with the adder at the bottom):
$$
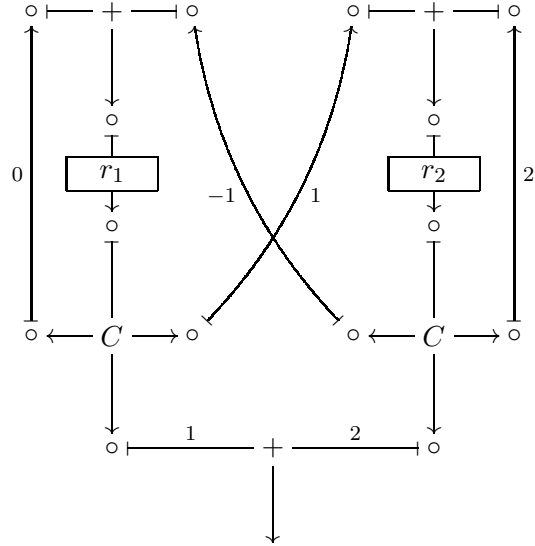
\xymatrix
@-3.0ex
@C+1.5ex
@R+4.0ex
{
\circ
\ar @{|-}[r]
&
+ \dto
&
\circ
\ar @{|-}[l]
& &
\circ \ar @{|-}[r]
&
+ \dto
&
\circ
\ar @{|-}[l]
\\
& \circ
\red{\ar @{|->}[d]|(.5)*+[F]{\;\;\;r_1 \;\;\; }}
& & & &
\circ
\red{\ar @{|->}[d]|(.5)*+[F]{ \;\;\;r_2 \;\;\;}}
&
\\
& \circ
\ar @{|-}[d]
& & & &
\circ
\ar @{|-}[d]
&
\\
\circ
\ar @{|->}[uuu]^0
&
C
\ar @{->}[l]
\ar @{->}[r]
\blue{\ar @{->}[d]}
&
\circ
\ar @/_1pc/ @{|->}[uuurr]_1
& &
\circ
\ar @/^1pc/ @{|->}[uuull]^{-1}
&
C
\ar @{->}[l]
\ar @{->}[r]
\blue{\ar @{->}[d]}
&
\circ
\ar @{|->}[uuu]_2
\\
& \red{\bf \circ} \ar @{|-}[rr]^-{1} &  & + \ar @{->}[d] & &
\red{\bf \circ} \ar @{|-}[ll]_-{2}  &
\\
& &  &   & &  &
}
$$
In the picture above, we
use $\circ$ to denote the composition of an output end with an
input end.
The circuit as a whole has no external input ends and
one external output end.
The heart of the circuit consists of two registers with initial
values $r_1$ and $r_2$. The outputs of the registers
are copied and:
\begin{itemize}
\item[(a)]
fed back to the input ends of the registers, via multipliers
whose values can be expressed by the following $2 \times 2$ matrix:
$$
M = \;
\left(
\begin{array}{cc}
0  & -1
\\
1 & 2
\end{array}
\right)
$$
This leads to new values of the registers given by
$$
\left(
\begin{array}{cc}
0  & -1
\\
1 & 2
\end{array}
\right)
\left(
\begin{array}{c}
r_1
\\
r_2
\end{array}
\right)
\, = \;\;
\left(
\begin{array}{c}
- r_2
\\
r_1 + 2r_2
\end{array}
\right)
$$
\item[(b)]
At the same time, the outputs of the registers are fed forward
into an adder combined with multipliers whose values
are given by the following matrix:
$$
N = \left(
\begin{array}{cc}
1 & 2
\end{array}
\right)
$$
This leads to a (first) output value given by
$$
\left(
\begin{array}{cc}
1 & 2
\end{array}
\right)
\left(
\begin{array}{c}
r_1
\\
r_2
\end{array}
\right)
= \;
r_1 + 2r_2
$$
\end{itemize}
We call this a circuit in {\em canonical form}.
More generally, we have the following definition.

\begin{definition}
\label{canonical form}
{\rm
We say that a stream circuit is in
{\em canonical form} if it has no input ends
and one output end; consists of $n \geq 1$
registers with feedback lines
given by an $n\times n$ matrix; and has
feedforward lines
given by an $1 \times n$ matrix leading via
an $n$-to-1 adder to a single output end.
}
\qed
\end{definition}

\medskip
\noindent
In the description of the example circuit above,
(a) and (b) together describe
one single atomic computation step of the circuit.
Next we describe the stream behaviour of our example
canonical circuit.
As we shall see,  the output end of a
canonical stream circuit produces precisely one
(rational) stream. In order to compute this output stream,
we first give names ($\sigma$ and $\tau$) to the streams that
will occur at the output ends of the two registers.
Then we apply the stream equations for each of the
basic gates in the circuit, leading to:
$$
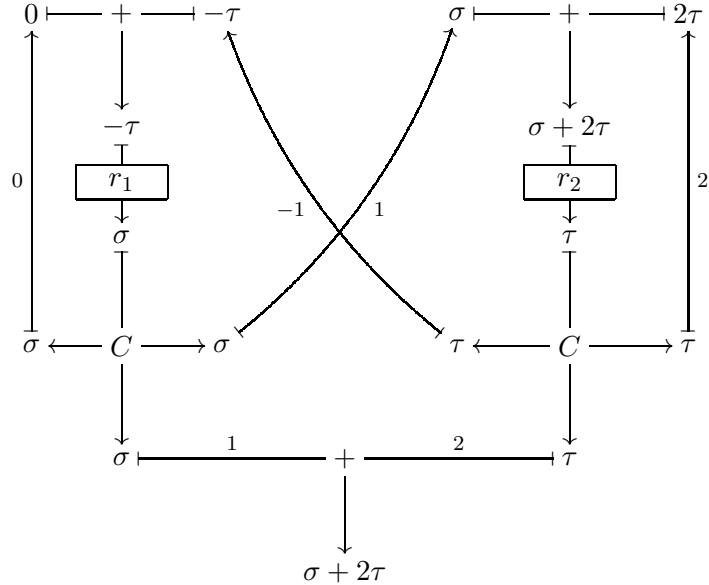
\xymatrix
@-3.0ex
@C+1.5ex
@R+4.0ex
{
0
\ar @{|-}[r]
&
+ \dto
&
- \tau
\ar @{|-}[l]
& &
\sigma \ar @{|-}[r]
&
+ \dto
&
2 \tau
\ar @{|-}[l]
\\
& - \tau
\red{\ar @{|->}[d]|(.5)*+[F]{\;\;\;r_1 \;\;\; }}
& & & &
\sigma + 2 \tau
\red{\ar @{|->}[d]|(.5)*+[F]{ \;\;\;r_2 \;\;\;}}
&
\\
& \sigma
\ar @{|-}[d]
& & & &
\tau
\ar @{|-}[d]
&
\\
\sigma
\ar @{|->}[uuu]^0
&
C
\ar @{->}[l]
\ar @{->}[r]
\blue{\ar @{->}[d]}
&
\sigma
\ar @/_1pc/ @{|->}[uuurr]_1
& &
\tau
\ar @/^1pc/ @{|->}[uuull]^{-1}
&
C
\ar @{->}[l]
\ar @{->}[r]
\blue{\ar @{->}[d]}
&
\tau
\ar @{|->}[uuu]_2
\\
& \red{\bf \sigma } \ar @{|-}[rr]^-{1} &  & + \ar @{->}[d] & &
\red{\bf \tau} \ar @{|-}[ll]_-{2}  &
\\
& &  & \sigma + 2\tau & &  &
}
$$
Applying the register law (\ref{register law})
to our two registers then leads to the following two equations
(writing $r$ for $[r]$ as usual):
\begin{eqnarray*}
\sigma
& = &
r_1 + (X \times -\tau)
\\
\tau
& = &
r_2 + (X \times (\sigma + 2\tau))
\\
\end{eqnarray*}
or, equivalently, in matrix notation:
$$
\left(
\begin{array}{c}
\sigma
\\
\tau
\end{array}
\right)
\, = \;\;
\left(
\begin{array}{c}
r_1
\\
r_2
\end{array}
\right)
\, + \;\; X \times \,
\left(
\begin{array}{cc}
0  & -1
\\
1 & 2
\end{array}
\right)
\; \times \;
\left(
\begin{array}{c}
\sigma
\\
\tau
\end{array}
\right)
$$
whence
$$
\left(
\begin{array}{cc}
1  & X
\\
-X \;\; & \; 1-2X
\end{array}
\right)
\; \times \;
\left(
\begin{array}{c}
\sigma
\\
\tau
\end{array}
\right)
\, = \;\;
\left(
\begin{array}{c}
r_1
\\
r_2
\end{array}
\right)
$$
This leads to the following values for $\sigma$ and $\tau$:
\begin{eqnarray*}
\left(
\begin{array}{c}
\sigma
\\
\tau
\end{array}
\right)
& = & \;
\left(
\begin{array}{cc}
1  & X
\\
-X \;\; & \; 1-2X
\end{array}
\right)^{-1}
\; \times \;
\left(
\begin{array}{c}
r_1
\\
r_2
\end{array}
\right)
\end{eqnarray*}
(recall that this inverse matrix was computed in
Example \ref{example linear system of differential equations}).
As a consequence the output stream of the circuit
is given by
\begin{eqnarray}
\label{stream behaviour of circuit}
\sigma + 2 \tau
& = &
\left(
\begin{array}{cc}
1 \;  & \; 2
\end{array}
\right)
\; \times \;
\left(
\begin{array}{cc}
1  & X
\\
-X \;\; & \; 1-2X
\end{array}
\right)^{-1}
\; \times \;
\left(
\begin{array}{c}
r_1
\\
r_2
\end{array}
\right)
\end{eqnarray}

We saw that the above circuit is fully determined
by the two matrices $M$ and $N$ containing the values of the
(feedback and feedforward) multipliers.
As such, the circuit corresponds precisely to a linear
system $(k^2 , \, <H,G> )$ with
$G: k^2 \to k^2$ and $H: k^2 \to k$
given by
$$
G(r_1,r_2) = \; M \times
\left(
\begin{array}{c}
r_1
\\
r_2
\end{array}
\right)
\;\;\;\;\;
H(r_1,r_2) = \; N \times
\left(
\begin{array}{c}
r_1
\\
r_2
\end{array}
\right)
$$
A state of this linear system corresponds to the
contents of the two registers of the circuit;
$G(r_1,r_2)$
corresponds to the feedback multiplication with the matrix
$M$; and the output given by $H(r_1,r_2)$ corresponds
to the feed-forward multiplication with the matrix $N$.
Note that the stream behaviour of our circuit as described
above corresponds precisely
with the (final) behaviour of the corresponding linear systems,
as given by (the proof of)
Theorem \ref{characterising finite dimensional linear O-systems}
in Section \ref{Linear representations}.
This follows from the fact that
identity (\ref{stream behaviour of circuit}) equals
\begin{eqnarray*}
\sigma + 2 \tau
& = &
\left(
\begin{array}{cc}
1 \;  & \; 2
\end{array}
\right)
\; \times \;
\left(
\begin{array}{cc}
1  & X
\\
-X \;\; & \; 1-2X
\end{array}
\right)^{-1}
\; \times \;
\left(
\begin{array}{c}
r_1
\\
r_2
\end{array}
\right)
\\
& = &
M_{[H]} \times \, M_{\tilde{G}} \, \times
\left(
\begin{array}{c}
r_1
\\
r_2
\end{array}
\right)
\end{eqnarray*}
with $H$ and $G$ as defined above.

Summarizing, we have presented an example of
a canonical stream circuit and shown that how to compute
the (rational) stream that it produces at its output end.
Then we observed that such
a canonical stream circuit corresponds precisely to a finite dimensional
linear system via its two matrices of feedback and feedforward
multipliers. Moreover, the stream behaviour of the circuit
coincides with that of the linear system.

In conclusion of this section, we note that one can construct,
conversely, from any finite dimensional linear system
$(V,<H,G>)$
a corresponding canonical stream circuit
with exactly the same stream behaviour:
the dimension of $V$ determines the number of registers;
the matrix corresponding to $G$ determines the values of
the feedback multipliers; and the matrix corresponding
to $H$ determines the values of the feedforward lines.

All in all, we have proved the following.

\begin{theorem}
\label{equivalence linear systems and stream circuits}
{\rm
Let $k$ be a field. For $\sigma \in k^\omega$,
the following are equivalent:
\begin{itemize}
\item[(1)]
The stream $\sigma$ is representable by a linear system of finite
dimension.
\item[(2)]
The stream $\sigma$ is computable by a finite stream circuit.
\end{itemize}
}
\qed
\end{theorem}

\section{Weighted stream automata}
\label{Weighted stream automata}

We saw that rational streams are ``finite memory'':
they can be computed by stream circuits with finitely
many registers. In this section, we show they are
also ``finite state'': they can be computed
by finite so-called weighted stream automata.

A \emph{weighted stream automaton} with values in a field $k$
is a pair $(Q,<o,t>)$
consisting of a set $Q$ of states, together with an output
function $o:Q \to k$ and a transition
function $t:Q \to (Q \to k)$. The output function $o$ assigns to
each state $q \in Q$ a value $o(q) \in k$ called the output
of $q$. The transition function $t$ assigns to
each state $q \in Q$ a function $t(q): Q \to k$,
which specifies for any state $q' \in Q$ a value
$t(q)(q') \in k$. This number can be thought of as
the \emph{weight} with which the
transition from $q$ to $q'$ occurs.
(There are various possible interpretations of this
notion of weight, such as the cost, multiplicity,
duration etc. of the transition.)
We will use the following notation:
\[
q \overto{r} q'
\;\;\equiv \;\; t(q)(q') = r \, , \;\;
\;\;\;\;\;\;
q \stackrel{r}{\Rightarrow}
\;\;\equiv \;\; o(q) = r
\]

Weighted stream automata represent streams in $k^\omega$ in the
following manner.

\begin{definition}
\label{definition S}
For a state $q \in Q$ of a weighted stream automaton
$(Q,<o,t>)$
we define a stream
$S(q)$, for all $k \geq 0$, by
\begin{eqnarray*}
\lefteqn{S(q)(k) =}
\\
\nonumber
& & \sum \, \{ \;
l_0 \times l_1 \times \cdots \times  l_{k-1} \times l \mid \;
\exists \, q_0, \ldots , q_k \,: \;
q= q_0 \overto{l_0} q_1 \overto{l_1} \cdots
\overto{l_{k-1} } q_k \stackrel{l}{\Rightarrow} \; \}
\end{eqnarray*}
(Here $\times$ denotes multiplication in the field $k$.)
\end{definition}

\medskip
\noindent
So the $kth$ value of the stream $S(q)$ is obtained by
considering all transition paths of length $k$ starting
in the state $q$; multiplying for each such path the
labels of all transitions; and adding up the resulting values
for all paths.
We say that the stream $S(q)$ is
{\it represented} by the state $q$.

\begin{example}
\label{example of weighted automaton}
Consider the following example of a weighted automaton:
$$
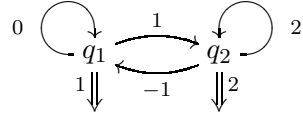
\xymatrix
@C+1.5ex
@R-2.0ex
{
q_1 \arrow @/^1.5ex/[r]^-{1}
\tolu^{\mathstrut 0\ }
\ddouble_(.45){1} |> \Tip
&
q_2 \arrow @/^1.5ex/[l]^-{-1}
\toru_{\mathstrut\ 2}
\ddouble^(.45){2} |> \Tip
\\
&
}
$$
Computing the streams $S(q_1)$ and $S(q_2)$
according to Definition \ref{definition S} above  gives
\[
S(q_1) = (1,2,3, \ldots \, ) \, ,
\;\;\;\;\;\;
S(q_2) = (2,3,4, \ldots \, )
\]
\qed
\end{example}

\medskip
\noindent
We can represent all information contained in the definition of
weighted stream automata by two matrices, in very much the
same way as we could define stream \emph{circuits} by two matrices as well.
To this end, we define
for a weighted stream automaton $(Q,<o,t>)$, with states
$\{ q_1 , \ldots , q_n \}$,
an output matrix $L$ and a transition matrix $K$ as follows:
\[
L_{i} = o(q_i) , \;\;\;\;\;
K_{ij} = t(q_i)(q_j)
\]
Now we can compute the streams represented
by the states of a weighted automaton directly in terms of these matrices.
Illustrating this for the example automaton above, we have
\[
L= \; \left(
\begin{array}{c}
1
\\
2
\end{array}
\right)
\;\;\;\;\;\;\;
K = \; \left(
\begin{array}{cc}
0 & 1
\\
\; -1 \; & 2\end{array}
\right)
\]
Applying Theorem \ref{fundamental theorem} to the (vector of) streams
\[
\sigma = S(q_1) , \;\;\;\;
\tau = S(q_2)
\]
we obtain
\begin{eqnarray*}
\left(
\begin{array}{c}
\sigma \\
\tau
\end{array}
\right)
& = &
\left(
\begin{array}{c}
\sigma (0) \\
\tau (0)
\end{array}
\right)
\;
+
\;\;
X \times
\left(
\begin{array}{c}
\sigma ' \\
\tau '
\end{array}
\right)
\end{eqnarray*}
Note that it follows from Definition \ref{definition S} that
\[
\left(
\begin{array}{c}
\sigma (0) \\
\tau (0)
\end{array}
\right)
=
\; L
\]
and
\[
\left(
\begin{array}{c}
\sigma ' \\
\tau '
\end{array}
\right)
= \;
K \times \,
\left(
\begin{array}{c}
\sigma \\
\tau
\end{array}
\right)
\]
As a consequence, we find
\begin{eqnarray*}
\left(
\begin{array}{c}
\sigma \\
\tau
\end{array}
\right)
& = &
\left(
\begin{array}{c}
1 \\
2
\end{array}
\right)
\;
+
\;\;
X \times \,
\left(
\begin{array}{cc}
0 & 1
\\
\; -1 \; & 2\end{array}
\right)
\times \,
\left(
\begin{array}{c}
\sigma \\
\tau
\end{array}
\right)
\end{eqnarray*}
which leads to
\begin{eqnarray*}
\left(
\begin{array}{c}
\sigma  \\
\tau
\end{array}
\right)
& = &
\left(
\begin{array}{cc}
1 & -X
\\
\; X \; & 1-2X
\end{array}
\right)^{-1}
\;
\times
\;
\left(
\begin{array}{c}
1
\\
2
\end{array}
\right)
\\
& = &
\frac{1}{(1-X)^2} \, \times \, \left(
\begin{array}{cc}
1-2X \; \; & X
\\
-X \; \; & 1
\\
\end{array}
\right)
\;
\times
\;
\left(
\begin{array}{c}
1
\\
2
\end{array}
\right)
\end{eqnarray*}
It follows that
\[
S(q_1) = \, \sigma = \, \frac{1}{(1-X)^2} \, , \;\;\;\;\;
S(q_2) = \, \tau = \,
\frac{2-X}{(1-X)^2}
\]
showing that the streams represented by our weighted automaton are
rational. All of the above generalises directly to arbitrary
weighted automata and so we have proved one half of the following
theorem.

\begin{theorem}
\label{weighted automata represent rational streams}
A stream $\sigma \in k^\omega$ is rational
iff it can be represented by a state $q \in Q$
of a finite weighted stream automata $(Q,<o,t>)$
with values in $k$.
\end{theorem}

\begin{proof}
The implication from right to left follows from the above.
For the converse, consider a rational stream $\sigma \in k^\omega$.
It follows from Theorem \ref{first equivalence theorem}
that $\sigma$ is representable by a linear system
of finite dimension $(k^n , <H,F>)$ with output in $k$.
Without loss of generality we can assume that $\sigma$ is represented
by the vector $(1, 0, \ldots, 0) \in k^n$.
We define $Q = \{ q_1 , \ldots , q_n \}$ by
\[
q_1 = (1,0, \ldots ,  0) , \; \ldots \; , \;
q_n = (0, \ldots ,  0 , 1)
\]
Next we define a weighted stream automaton $(Q,<o,t>)$
by putting, for all $1 \leq i,j\leq n$,
\[
o(q_i) = \, H_i , \;\;\;\;\;\;
t(q_i)(q_j) = \, F^T_{ij} = \, F_{ji}
\]
It follows that $\sigma = S(q_1)$, that is,
$\sigma$ is represented by the state
$q_1$ in $(Q,<o,t>)$.
\end{proof}

\section{Summary and discussion}
\label{Summarizing the results}

All in all, we have proved the following.

\begin{theorem}
\label{general equivalence theorem}
{\rm
Let $k$ be a field. For a stream $\sigma \in k^\omega$,
the following are equivalent:
\begin{itemize}
\item[(1)]
The stream $\sigma$ is rational: $\sigma = \rho/\tau$
for polynomial streams $\rho$ and $\tau$ (with $\tau(0) \neq 0$).
\item[(2)]
The stream $\sigma$ is representable by a linear system
of finite dimension.
\item[(3)]
The subsystem $Z_\sigma \subseteq (k^\omega,<i,d>)$
generated by $\sigma$ has finite dimension.
\item[(4)]
The stream $\sigma$ is computable by a finite stream circuit.
\item[(5)]
The stream $\sigma$ is representable by
a finite weighted stream automaton.
\end{itemize}
\qed
}
\end{theorem}

\noindent
We mention a few examples of the
many interesting questions and directions that remain to be explored.
Streams over a {\em finite} field enjoy many special properties.
A special example is the family of bitstreams, which consist of 0's and 1's.
The interplay between coalgebraic techniques and various algebraic structures on bitstreams, such as
the Boolean and the 2-adic operators, deserves further study, which may also be relevant
for the construction and analysis of digital circuits; see \cite{Rut05,HCR06} for some
preliminary results. There is also much and interesting life beyond rationality.
For instance, it would be worthwhile to try and apply coinductive techniques
to the study of so-called {\em automatic} sequences, see for instance
\cite{AS03}. Another example is the combined use of
linear systems theory and coalgebra in the
world of hybrid systems, where discrete time and continuous time phenomena
occur simultaneously. The relationship between
rational streams and $\omega$-regular infinite words from formal language theory
is yet another subject that deserves further study.


\bigskip
\noindent
{\bf Acknowledgments\/}:
This paper, as well as an earlier version of it, has been reviewed by
anonymous referees. I am very grateful for the many constructive comments
these referees have made. They have improved both the presentation of the paper and
my understanding of its contents.

\appendix
\section{}
A  {\em semi-ring\/}
$A=(A,\,+,\,\cdot,\, 0,\,1)$
is a set $A$ with a commutative operation of
addition $c + d$;  a (generally non-commutative)
operation of multiplication $ c \cdot d$ with
$c \cdot (d + e ) = (c \cdot d) + (c \cdot e)$ and
$(d + e ) \cdot c = (d \cdot c) + (e \cdot c)$;
and with neutral elements $0$ and $1$ such that
$c + 0 = c$, $1 \cdot c = c \cdot 1 = c$
and $c \cdot 0 = 0 \cdot c = 0$.
If every $c \in A$ moreover has an additive inverse $-c$
(with $c + (-c) = 0$) then $A$ is a \emph{ring}.
If moreover multiplication is commutative and every
(non-zero) element $c \in A$ has a multiplicative inverse
$c^{-1}$ (with $c \cdot c^{-1} = 1$) then $A$ is a {\em field}.

\end{document}